# Interaction between Winds from Weak-lined T Tauri Stars with Exoplanetary Magnetospheres

Yasmmin F. Tamburus, Natália F. S. Andrade, Guilherme R. C. Sampaio, and Vera Jatenco-Pereira
Instituto de Astronomia, Geofísica e Ciências Atmosféricas da Universidade de São Paulo (IAG/USP), Rua do Matão 1226, Cidade Universitária, São Paulo, 05508-090, Brazil; ytamburus@gmail.com

## Abstract

T Tauri stars, in more advanced stages of evolution, during the final accretion phase of stellar formation, exhibit intense stellar winds and surface magnetic fields with intensities around a kilogauss. With the growing interest in the search for rocky exoplanets with Earth-like dimensions, it is essential to deepen our understanding of the interaction between stellar winds and planetary magnetospheres. We investigated the interaction between stellar winds from 46 weak-lined T Tauri stars (WTTSs) and the magnetospheric protection of Earth-like planets located within their habitable zones. We employ two distinct stellar wind models, nonmagnetized and magnetized with both constant and resonant Alfvén wave damping, to evaluate the pressure balance between the stellar wind and the planetary magnetic field. Our results show that the strong wind dynamic and magnetic pressures characteristic of WTTSs lead to systematically compressed planetary magnetospheres, significantly smaller than that of the present-day Earth. The analysis further indicates that planetary magnetospheric sizes increase with stellar age, following the decay of stellar magnetic activity, in agreement with previous findings for solar-type stars.

*Unified Astronomy Thesaurus concepts:* Stellar winds (1636); Protostars (1302); Magnetohydrodynamics (1964); Magnetic fields (994); Star-planet interactions (2177)

## 1. Introduction

Weak-lined T Tauri stars (WTTSs) are young (<10 Myr) and low-mass ($M_\star \leqslant 2\,M_\odot$) pre-main-sequence (PMS) stars that exhibit strong chromospheric activity and emission lines. They are defined by the weakness or absence of H$\alpha$ emission in their spectra, indicating a lack of active accretion. WTTSs also do not possess optically thick circumstellar disks, suggesting that they represent a more evolved phase in which the disk has become optically thin or has already dispersed (A. Ghez 1993). These stars are primarily identified through their intense X-ray emission (R. Neuhäuser et al. 1995), and are thought to drive powerful ionized winds with mass-loss rates on the order of $10^{-12}$–$10^{-9}\,M_\odot\,\mathrm{yr}^{-1}$ (A. A. Vidotto & J. F. Donati 2017)—several orders of magnitude higher than that of the solar wind. At these early evolutionary stages, convective processes generate strong magnetic fields, as evidenced by Zeeman broadening and polarimetry measurements (J.-F. Donati & J. Landstreet 2009; C. A. Hill et al. 2019; F. Pérez Paolino et al. 2025).

Young, magnetically active stars generate stronger, denser stellar winds than older stars, due to their rapid rotation and enhanced magnetic activity, which can significantly influence the plasma environment of nearby planets (A. A. Vidotto et al. 2013; V. See et al. 2014; D. Ó Fionnagáin & A. A. Vidotto 2018). The interaction between the outflowing stellar wind and a planet's intrinsic magnetic field determines the size of the planet's magnetosphere, often quantified by the magnetopause standoff distance ($R_s$) where the dynamic pressure of the wind is balanced by the magnetic pressure of the planet (G. L. Siscoe & D. G. Sibeck 1980). Under intense stellar-wind conditions, this interaction shapes the surrounding plasma environment, determining the structure and how far the planetary magnetic field can stand off from the incoming flow (O. Cohen et al. 2014). Observationally and in modeling studies, such stellar wind–planet interactions produce structures such as compressed magnetospheres and shock boundaries that depend on the wind properties and the magnetic characteristics of the planet (A. Canet et al. 2024). During early stages of stellar evolution, when winds are stronger and planetary magnetic fields are still evolving, quantifying the stellar-wind conditions and deriving the corresponding magnetopause distances is a key step in understanding the nature of star–planet interactions in young systems (D. Kubyshkina et al. 2022).

Planetary magnetospheres are frequently discussed in the context of atmospheric evolution and long-term surface conditions, as they regulate the interaction between a planet and the surrounding stellar-wind plasma. However, the relationship between magnetospheric size, atmospheric retention, and habitability is complex and remains poorly constrained (J. A. Tarduno et al. 2014; J. Rodríguez-Mozos & A. Moya 2017). In contrast, planets with weak or absent magnetic fields are more directly exposed to stellar-wind particles, potentially altering atmospheric chemistry (H. Lammer et al. 2008; G. Hazra 2025). As a result, there is currently no consensus on whether a strong planetary magnetic field systematically promotes atmospheric retention or enhances habitability (H. Gunell et al. 2018; R. Ramstad & S. Barabash 2021).

V. See et al. (2014) investigated the magnetospheric protection of planets against stellar winds from solar-like stars for two different stellar-wind models, those of E. N. Parker (1958) and S. R. Cranmer & S. H. Saar (2011), considering two key criteria in the search for extrasolar Earth analogs: the exoplanet must reside within the habitable zone (HZ), and it must possess a magnetosphere large enough to shield its atmosphere. They evaluated the ability of exoplanets to sustain magnetospheres comparable in size to the present-day Earth and the early Earth during the Paleoarchean era. This geological period, ∼3490–4030 Myr ago, marks the first evidence of macroscopic fossil life (F. M. Gradstein et al. 2012), where J. A. Tarduno et al. (2010) estimated the terrestrial magnetic field strength to be 50% that of the present-day field and suggested steady-state magnetopause standoff distances ($R_s$) of ⩾5 $R_?$. V. See et al. (2014) calculated the Paleoarchean magnetospheric size for the E. N. Parker (1958) and S. R. Cranmer & S. H. Saar (2011) stellar-wind models and obtained values ranging from 5 to 7 $R_?$. R. K. Bono et al. (2022) characterized the time-varying paleomagnetic field using different methods, and for most of the Precambrian period (∼540–4560 Myr ago, which includes the Paleoarchean), they calculated an average standoff distance of ∼6 $R_?$, which is about 60% of the present-day value.

In this paper, we investigate the stellar-wind environment of WTTSs and its implications for the magnetopause standoff distance of an Earth-like planet located at the HZ distance of each star. Using X-ray luminosity ($L_X$) measurements from the XMM-Newton Extended Survey of the Taurus molecular cloud (M. Güdel et al. 2007), we analyze a sample of 46 WTTSs with ages ranging from approximately 0.5–10.5 Myr. Although a single star has an age limit above 10 Myr (see Table A1), it was still classified as a WTTS. Of these, 12 had duplicate records, the result of discrepancies in the measurement of X-ray luminosity. For this reason, both luminosity values were taken into account in the calculations. For each star, we estimate the location of the HZ based on stellar parameters and apply two stellar-wind models, those of E. N. Parker (1958) and V. Jatenco-Pereira & R. Opher (1989a, hereafter JPO89), to derive the stellar-wind properties at that distance. These wind conditions, combined with assumed values for the planetary magnetic field appropriate for early evolutionary stages, are then used to estimate the magnetopause standoff distance. Our results are then compared with those of V. See et al. (2014). We also examine the relation between magnetospheric size and stellar age, based on the premise that younger stars exhibit stronger magnetic activity (S. Bellotti et al. 2025). As stars evolve and their magnetic activity declines, the magnetospheric protection provided by a planet with a fixed intrinsic magnetic field is expected to increase over time.

The paper is structured as follows. Section 2 presents the stellar-wind models, as well as stellar-wind pressures and chromospheric activity models. Section 3 focuses on the HZ model and Earth-like planets' magnetospheric sizes. Section 4 discusses the results of the models introduced in the previous sections, including comparisons with the findings of other studies. Finally, Section 5 summarizes the main conclusions drawn from the discussion.

## 2. Stellar-wind Models

### 2.1. E. N. Parker (1958) Wind Model

The model proposed by E. N. Parker (1958) examines that the gas frequently flows radially in all directions from the Sun at velocities ranging from approximately 200–800 km s$^{-1}$, depending on latitude, as observed by the Ulysses spacecraft (E. Echer et al. 2022). In coronal mass ejection events and flares, this velocity increases. E. N. Parker (1958) developed his solar wind model under the assumption that the magnetic field **B** and the system's rotation can be neglected, adopting a coronal temperature of approximately ∼10$^6$ K. Also, he considered a stationary ($\partial_t = 0$), isothermal, and spherically symmetric stellar-wind model. Starting from the momentum equation given by

$$\rho(\boldsymbol{u} \cdot \nabla)\boldsymbol{u} = -\nabla P + \rho \mathbf{g}, \quad (1)$$

where $\rho$ is the wind density, $\boldsymbol{u}$ is the expansion velocity, which is purely radial $\boldsymbol{u} = u_r \hat{\boldsymbol{r}} = u$, $\mathbf{g} = \nabla \psi$, with $\psi = -GM_\odot/r$ being the gravitational potential, and $P$ is the pressure. Besides, the continuity equation is

$$\nabla \cdot (\rho \boldsymbol{u}) = 0. \quad (2)$$

Since the system is spherical and stationary, Equation (2) takes the form $\rho = \text{constant}/ur^2$. Adopting the isothermal speed of sound as $c_s^2 = P/\rho$, the velocity profile for E. N. Parker (1958)'s model is given by

$$\left(\frac{u}{c_s}\right)^2 - \ln\left(\frac{u}{c_s}\right)^2 - 4\ln\left(\frac{r}{r_c}\right) - 4\frac{r_c}{r} + 3 = 0, \quad (3)$$

where $r_c$ is the critical radius (sonic point), with solutions $u = c_s$, $r = r_c$, and

$$r_c = \frac{GM_\star}{2c_s^2}. \quad (4)$$

This solution assumes an initially subsonic wind for $r < r_c$, which becomes supersonic after passing the critical point, i.e., for $r > r_c$.

#### 2.1.1. Wind Pressure in the E. N. Parker (1958) Model

The density profile is based on the mass conservation from Equation (2), assuming a density normalized ($\tilde{\rho}$) to the initial density $\rho_0$ (V. See et al. 2014):

$$\tilde{\rho} = \frac{\rho}{\rho_0} = \frac{u_0 r_0^2}{ur^2}, \quad (5)$$

where $r_0$ is the stellar radius and $u_0$ is the initial wind velocity. To estimate the wind pressure, we adopt the density at the closed corona, even though the E. N. Parker (1958) model does not consider the presence of a magnetic field. This density can be estimated based on chromospheric activity using a relation for the X-ray luminosity parameterized by solar parameters:

$$\rho_{c\star} = \rho_{c\odot} \left[\left(\frac{r_\odot}{r_\star}\right)^3 \left(\frac{L_{X\star}}{L_{X\odot}}\right)\right]^{1/2}, \quad (6)$$

where $\rho_{c\odot} = m\tilde{n}_{c\odot}$, with $\tilde{n}_{c\odot} = 10^8$ cm$^{-3}$ corresponding to the solar coronal number density (M. Guhathakurta et al. 1996), and $L_{X\odot} = 2.24 \times 10^{27}$ erg s$^{-1}$ is the Sun's X-ray luminosity (P. G. Judge et al. 2003). We stress that the relation adopted to link the coronal base density to the disk-integrated X-ray luminosity should be interpreted as an order-of-magnitude scaling rather than a strict observational constraint. Solar and stellar-wind studies indicate that wind mass flux shows only a weak correlation with X-ray emission and can vary over orders of magnitude, as it is primarily controlled by open magnetic flux rather than global coronal activity (O. Cohen 2011; C. Johnstone et al. 2015). Therefore, Equation (6) is employed here as a convenient prescription to estimate wind densities, and its limitations are explicitly taken into account by exploring a broad range of values in the uncertainty analysis. From Equations (1), (5), and (6), we obtain the ram pressure of the stellar wind:

$$P_{\rm ram} = \rho u^2 = \rho_{c\star} f \tilde{\rho} u^2, \quad (7)$$

where $f = 0.1$ (M. Guhathakurta et al. 1996) is a factor that considers the difference between the densities of coronal holes and the closed corona. For the E. N. Parker (1958) model, we assume the stellar-wind pressure $P_{\rm sw} = P_{\rm ram}$.

### 2.2. Alfvén-driven Wind Model

Coronal holes are known to be the source of high-speed solar wind streams at Earth's orbit (H. Peter & P. Judge 1999). Their characteristics, including open magnetic field lines and lower coronal densities, indicate enhanced solar wind acceleration near the Sun's surface (R. H. Munro & B. V. Jackson 1977). T. E. Holzer & E. Leer (1980) and V. Jatenco-Pereira & R. Opher (1989b) observed that the geometry of magnetic field funnels within coronal holes could significantly impact both the mass flux and wind velocity. R. Esser et al. (2005) demonstrated that models incorporating highly divergent magnetic funnels provide a better explanation for the observed solar wind data. There is direct evidence of Alfvén waves in the solar wind (J. W. Belcher & L. Davis 1971; E. Echer et al. 2025).

Alfvén waves do not produce density perturbations. Therefore, in a medium with density $\rho$, their propagation speed is given by

$$V_{\rm A} = \frac{B}{\sqrt{4\pi\rho}}. \quad (8)$$

In a homogeneous plasma of electrons and ions, the propagation frequency follows the dispersion relation:

$$\omega = \frac{1}{\sqrt{1 + (V_{\rm A}/c)^2}} k |\cos\theta| V_{\rm A}, \quad (9)$$

where $\theta$ is the angle between the magnetic field and the wave propagation direction $\boldsymbol{k}$. For $V_{\rm A} \ll c$ and waves propagating along the magnetic field lines ($\theta = 0$), the dispersion relation takes the simplified form $\omega = k V_{\rm A}$. Considering the presence of Alfvén waves in the solar wind, which is characterized by magnetic field perturbations without associated density fluctuations, JPO89 proposed that a flux of Alfvén waves serves as a mechanism for accelerating the wind of a protostar. We will apply this same model to WTTSs. The magnetic field geometry used is the same as that of a solar coronal hole (see Figure 1 from V. Jatenco-Pereira & R. Opher 1989b).

The regions of open magnetic field in the corona correspond to the coronal holes, as well as the source of high-speed flux (A. S. Krieger et al. 1973; J. T. Nolte et al. 1976). Coronal holes are observed on the Sun's surface, where the solid angle ($\Omega_0$) varies by approximately $\sim 7\Omega_0$ in the region between $r = R_\odot$ and $r = 3\,R_\odot$. For $r > 3\,R_\odot$, the coronal hole becomes essentially radial and follows the relation $F = \Omega/\Omega_0 = 7.26$ (R. H. Munro & B. V. Jackson 1977). N. P. M. Kuin & A. G. Hearn (1982) used a simple analytical form to describe the divergent geometry, where $A(r)$ is the cross section of the geometry at distance $r$. In our case:

$$A(r) = \begin{cases} A(r_0)\left(\dfrac{r}{r_0}\right)^S, & \text{for } r \leqslant r_{\rm T}, \\ A(r_0)\left(\dfrac{r_{\rm T}}{r_0}\right)^S \left(\dfrac{r}{r_{\rm T}}\right)^2, & \text{for } r > r_{\rm T}. \end{cases} \quad (10)$$

The geometry of the magnetic field, which transitions from divergent ($S > 2$) to radial ($S = 2$) at the transition radius ($r_{\rm T}$), is given by

$$F = \frac{\Omega}{\Omega_0} = \frac{A(r_T)/r_{\rm T}^2}{A(r_0)/r_0^2} = \left(\frac{r_{\rm T}}{r_0}\right)^{S-2}. \quad (11)$$

Adopting $F = 10$, then $r_{\rm T} = 10^{1/(S-2)} r_0$, where $r_0$ is the radius of the star, and $S$ is a parameter that determines the divergence of the geometry.

The fundamental equations that describe the Alfvén-wave-driven wind model are based on the mass, momentum, and energy conservation equations, with the energy equation being necessary when the temperature behavior along the wind is not known. In our case, an isothermal wind is assumed. Assuming the wind is one-dimensional, the mass continuity equation is given by Equation (2); in the geometrically spherical and stationary case, we have $\rho = \rho_0 u_0 A(r_0)/u A(r)$. From Equation (1), the momentum equation is

$$u\frac{du}{dr} = -\left(\frac{GM_\star}{r^2} + \frac{1}{\rho}\frac{dP}{dr} + \frac{1}{2\rho}\frac{d\epsilon}{dr}\right), \quad (12)$$

with (L. Hartmann & K. B. MacGregor 1980)

$$\epsilon = \epsilon_0 \frac{M_{\rm A0}}{M_{\rm A}}\left(\frac{1 + M_{\rm A0}}{1 + M_{\rm A}}\right)^2 \exp\left\{-\int_{r_0}^{r}\frac{1}{L}dr'\right\}, \quad (13)$$

where $\epsilon = \rho\langle\delta v^2\rangle$ is the energy density of the Alfvén wave, $\langle\delta v^2\rangle$ is the mean square amplitude velocity fluctuations, $M_{\rm A}$ is the Alfvénic Mach number, defined as $M_{\rm A} = u/V_{\rm A}$, and $L$ is the damping length. The factor $\epsilon_0$ can be calculated from the initial energy flux of the waves $\phi_{M0} = \epsilon_0 V_{\rm A0}(1 + 3M_0/2)$, where the energy flux is given by

$$\phi_M = \phi_{M0}\left(\frac{1 + 3M_{\rm A}/2}{1 + 3M_{\rm A0}/2}\right)\left(\frac{1 + M_{\rm A0}}{1 + M_{\rm A}}\right)^2 \\ \exp\left\{-\int_{r_0}^{r}\frac{1}{L}dr'\right\}. \quad (14)$$

The equation that describes the Alfvén-wave-driven wind model is given by (JPO89)

$$\frac{1}{u}\frac{du}{dr}\left[u^2 - V_{\rm T}^2 - \frac{1}{4}\left(\frac{1 + 3M_{\rm A}}{1 + 4M_{\rm A}}\right)\langle\delta v^2\rangle\right] \\ = \frac{Z}{r}\left[V_{\rm T}^2\left(1 - \frac{r}{V_{\rm T}}\frac{dV_{\rm T}}{dr}\right) + \frac{1}{4}\left(\frac{1 + 3M_{\rm A}}{1 + M_{\rm A}}\right)\langle\delta v^2\rangle\right. \\ \left. + \frac{1}{4}\frac{r}{L}\langle\delta v^2\rangle - \frac{v_{\rm esc}^2}{2Z}\right]. \quad (15)$$

Since we adopt an isothermal wind, we consider the thermal velocity equal to the sound speed, $V_{\rm T} = c_s$, so that $dV_{\rm T}/dr = 0$. The escape velocity is given by $v_{\rm esc} = \sqrt{2GM_\star/r_0}$, and $Z$ is

defined by

$$Z = \begin{cases} S, & \text{for } r \leqslant r_T, \\ 2, & \text{for } r > r_T. \end{cases} \quad (16)$$

For the JPO89 model, we use constant and resonant damping of the damping length, $L$, where resonant damping is described in Section 2.2.1.

### 2.2.1. Resonance Surface Absorption Damping

In the context of the solar atmosphere, J. A. Ionson (1978) observed that resonant damping effectively reproduces the heating of coronal loops. In the divergent geometry of an open magnetic field, we can have a layer thickness ($a$), where Alfvén waves with different frequencies $\omega_1$ and $\omega_2$ can propagate outside and inside the inhomogeneity, respectively. As noted by J. V. Hollweg (1987), the resonant layer extracts energy from the external surface wave, causing the wave to decay, and its energy reappears in the resonant layer. The resonant damping rate is given by (M. A. Lee & B. Roberts 1986)

$$\Gamma = \pi \bar{k} a \frac{\Delta}{4\bar{\omega}}, \quad (17)$$

where $\bar{k}$ is the average wavenumber of the surface wave ($\bar{k}a < <1$), $\bar{\omega}$ is a linear variation of the wave frequency, with $\bar{\omega}^2 = (\omega_2^2 + \omega_1^2)/2$, and $\Delta = (\omega_2^2 - \omega_1^2)/2$. In the regime where $\omega_2^2 >> \omega_1^2$:

$$\Gamma = \bar{\omega}\left(\frac{\pi k a}{4\sqrt{2}}\right). \quad (18)$$

The expression of the damping length is given by

$$L = \frac{V_A}{\Gamma}. \quad (19)$$

The following equations describe the divergent and radial geometries of magnetic field lines. For $r \leqslant r_T$ and $Z = S$:

$$\rho u \left(\frac{r}{r_0}\right)^S = \rho_0 u_0, \quad (20)$$

$$B = B_0 \left(\frac{r_0}{r}\right)^S, \quad (21)$$

$$L = L_0 \left(\frac{V_A}{V_{A0}}\right)^2 \left(\frac{r_0}{r}\right)^{S/2} (1 + M_A). \quad (22)$$

And for $r > r_T$ and $Z = 2$:

$$\rho u \left(\frac{r_T}{r_0}\right)^S \left(\frac{r}{r_T}\right)^2 = \rho_0 u_0, \quad (23)$$

$$B = B_0 \left(\frac{r_0}{r_T}\right)^S \left(\frac{r_T}{r}\right)^2, \quad (24)$$

$$L = L_0 \left(\frac{V_A}{V_{A0}}\right)^2 \left(\frac{r_0}{r_T}\right)^{S/2} \left(\frac{r_T}{r}\right) (1 + M_A). \quad (25)$$

### 2.2.2. Wind Pressure in the V. Jatenco-Pereira & R. Opher (1989a) Model

The JPO89 model assumes a magnetized stellar wind, and therefore we must consider the contributions of magnetic and ram pressures. Following V. See et al. (2014), we used a ram pressure given by

$$P_{\text{ram}} = \frac{\dot{M} v_{\text{esc}}}{4\pi r^2}, \quad (26)$$

where $\dot{M} = 4\pi u_0 \rho_0 r_0^2$ is the stellar wind's mass-loss rate. In addition, we calculated the magnetic pressure using

$$P_B = \frac{B^2}{8\pi}, \quad (27)$$

with the magnetic field calculated from Equations (21) and (24). For the JPO89 model, we assume the stellar-wind pressure $P_{\text{sw}} = P_{\text{ram}} + P_B$.

### 2.3. Chromospheric Activity

To calculate the chromospheric activity of WTTSs, which are nonaccreting PMS stars of spectral class III, we used the model proposed by B. Stelzer et al. (2013). This model relates two different stellar fluxes, where we chose the stellar X-ray emission with the H and K lines of Ca II through the following equation:

$$\log F_X = c_1 + c_2 \log F_{HK}, \quad (28)$$

with $c_1 = +1.04$ and $c_2 = +1.06$. We can calculate the chromospheric activity, $R_{HK}$, measured in the H and K lines of Ca II using the relation $R_{HK} = L_{HK}/L_{\text{bol}} = F_{HK}/\sigma T_{\text{eff}}$ (J. L. Linsky et al. 1979), as well as $R_X = L_{X\star}/L_{\text{bol}} = F_X/\sigma T_{\text{eff}}$ for X-rays. B. Stelzer et al. (2013) estimated $c_1$ and $c_2$ based on X-ray and Ca II K emissions; since we want a measurement of the H and K lines of Ca II, we add a correction of a factor of 2, where $F_{HK} = F_H + F_K = 2F_K$. Here, we consider Ca II H and Ca II K to have the same value.

## 3. Habitable Zone and Planetary Magnetosphere

Based on J. F. Kasting et al. (1993) and R. K. Kopparapu et al. (2013a), who analyzed an Earth-mass planet with an atmosphere dominated either by $H_2O$ (inner edge—moist greenhouse) or $CO_2$ (outer edge—maximum greenhouse), the boundaries of the HZ around F-, G-, K-, and M-type stars were established using an one dimensional, cloud-free radiative–convective climate model. In this method, the temperature of the planetary surface is specified in advance (ranging from 200 to 2200 K), and the model is used to compute the corresponding effective stellar flux ($S_{\text{eff}}$) required to sustain it, which provides a more reliable criterion for defining the HZ boundaries. $S_{\text{eff}}$ represents the solar constant required to maintain a given surface temperature and is calculated as the ratio between the outgoing infrared flux ($F_{\text{IR}}$) and the incident solar flux ($F_\odot$), both evaluated at the top of the atmosphere, $S_{\text{eff}} = F_{\text{IR}}/F_\odot$ (G. Siscoe & C.-K. Chen 1975):

$$S_{\text{eff}} = S_{\text{eff}\odot} + aT_\star + bT_\star^2 + cT_\star^3 + dT_\star^4, \quad (29)$$

where $T_\star = T_{\text{eff}} - 5780$ K, with $T_{\text{eff}}$ being the effective star temperature. The constants $S_{\text{eff}\odot}$, $a$, $b$, $c$, and $d$ are based on tabulated values of HZ boundaries (R. K. Kopparapu et al. 2013b). Finally, the radial distance of the HZ is calculated using the star luminosity by

$$r_{\rm HZ} = \left(\frac{L/L_\odot}{S_{\rm eff}}\right)^{1/2} {\rm au}. \quad (30)$$

The magnetosphere of a planet is shaped by the interaction between the stellar wind and the intrinsic magnetic field of the planet, and is governed by the balance between different forms of pressure in the surrounding plasma. The model of exoplanet magnetospheres, based on J. M. Grießmeier et al. (2004), considers that the magnetic field within the magnetosphere is represented as a superposition of the planet's intrinsic dipolar magnetic field and an external field generated by magnetopause currents. The overall geometry of the magnetosphere is approximated as a hemispherical shape on the dayside and a semi-infinite cylindrical tail on the nightside. The size of the magnetosphere, which is the standoff distance ($R_s$), is determined by the pressure balance of the stellar wind and the planetary magnetosphere:

$$R_{\rm s} = \left(\frac{\mu_0 f_0^2 \mathcal{M}^2}{8\pi^2 P_{\rm sw}}\right)^{1/6}, \quad (31)$$

where $\mu_0$ is the magnetic permeability, $\mathcal{M}$ is the planetary magnetic moment, and $f_0 = 1.16$ is a factor that takes into account the nonspherical shape of the magnetopause (H. Volland 1995).

## 4. Results and Discussion

Combining the models described above, we investigated the interaction between the stellar winds from WTTSs and the magnetospheres of Earth-like exoplanets. We placed a hypothetical Earth analog within each star's HZ and evaluated the pressure balance at the magnetopause as a proxy for the characteristic size of the planetary magnetosphere. The stellar-wind velocity profiles were calculated using the models proposed by E. N. Parker (1958) and JPO89. We then determined the extent of the HZ using the approach of R. K. Kopparapu et al. (2013a) and placed the planet midway between the moist greenhouse and maximum greenhouse limits, that is, at the center of the HZ. At this orbital distance, we evaluated the local stellar-wind velocity and pressure predicted by each wind model and computed the corresponding magnetopause standoff distance assuming pressure balance with a terrestrial-like planetary magnetic field (J. M. Grießmeier et al. 2004). We emphasize that the E. N. Parker (1958) wind is adopted here as a reference framework, providing a first-order baseline rather than a complete physical description of stellar winds. Observational and theoretical studies indicate that additional processes, such as wave-driven acceleration and coronal heating, may contribute to the acceleration and energetics of stellar winds (L. Ofman 2010). While one-dimensional wind models offer a practical approach under simplified assumptions, stellar-wind properties may also depend on physical effects not included in such models, such as magnetic field topology and multidimensional plasma dynamics (A. Vidotto et al. 2014; C. Garraffo et al. 2018). These effects are not explicitly modeled in this work and are therefore treated as additional sources of uncertainty in the interpretation of the results.

First, we analyzed the behavior of the X-ray to bolometric luminosity ratio as a function of stellar age, comparing our WTTSs sample with the solar-type stars from V. See et al. (2014). As shown in Figure 1, the data reveal an exponential decline of this ratio with increasing age. Nevertheless, the WTTSs clearly show enhanced $L_{\rm X}$ emission, consistent with their strong magnetic activity, and therefore are also expected to display elevated levels of chromospheric activity (B. Stelzer et al. 2003; H. M. Günther et al. 2010; E. Flaccomio et al. 2012). We did not adopt the chromospheric activity model of E. E. Mamajek & L. A. Hillenbrand (2008) as V. See et al. (2014), because our WTTS sample lies in the saturated regime ($\log R_{\rm X} > -4$), where the correlation between chromospheric and coronal activity breaks down and the model is no longer reliable.

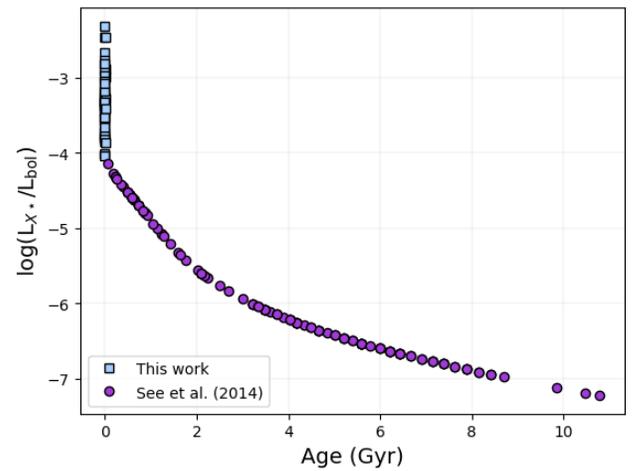

**Figure 1.** Ratio of X-ray to bolometric luminosity $\log(L_{\rm X\star}/L_{\rm bol})$ as a function of stellar age, comparing WTTS data with solar-type stars from V. See et al. (2014).

For the E. N. Parker (1958) wind model, we calculated the wind speed profile from the stellar radius $r_0$ up to 3.5 au ($\sim 566\, r_0$). We also have adopted a temperature of 2.1 MK, because it reproduces the solar wind parameters on Earth (V. See et al. 2014). WTTSs are expected to have coronal temperatures in the range of millions of kelvin (A. Telleschi et al. 2007; E. L. Jensen et al. 2009; W. W. Golay et al. 2023). For the JPO89 model, we estimate profiles from $r_0$ to $300\, r_0$, and for all stars we assume an initial magnetic field of $B_0 = 1$ kG (A. Vidotto et al. 2009; T. Carroll et al. 2012) and a base wind density of $\rho_0 = 1 \times 10^{-11}$ g cm$^{-3}$ (A. Vidotto et al. 2010). In the JPO89 model, we adopted a coronal temperature of $2.1 \times 10^6$ K for only a subset of stars, while for the remaining stars we used a lower temperature of $2.1 \times 10^5$ K. This distinction was necessary because, given the initial parameters used in Equation (15), not all stars can sustain coronal temperatures on the order of $10^6$ K. For some cases, such high temperatures prevent the wind solution from transitioning from subsonic to supersonic regimes, which is a critical condition for a physically viable stellar-wind solution. We expect the values of the terminal velocity ($u_\infty$) to be $\sim 310$ km s$^{-1}$ for an isothermal wind with $5 \times 10^5$ K, and $\sim 760$ km s$^{-1}$ for $2 \times 10^6$ K (S. Preusse et al. 2005). Furthermore, in the JPO89 model, we varied the free parameters $S$, $L_0$, and $\phi_{M0}$ to try to reproduce the expected terminal velocity. Figure 2 presents, as an example, the velocity profiles obtained for the HBC 359 star. Notably, the sonic point coincides with the initial velocity in the E. N. Parker (1958) model for this particular star, indicating that the stellar wind becomes supersonic from that point onward. The sonic point does not coincide with the stellar radius for all stars. Additionally, for the HBC 359 star, we adopt the following parameters. For resonant damping, we use $S = 4.5$, $L_0 = 4.0\,r_0$, and $\phi_{M0} = 3.36 \times 10^9$ erg cm$^{-2}$ s$^{-1}$, resulting in $u_\infty = 373.3$ km s$^{-1}$. For constant damping, we adopt $S = 3.5$, $L_0 = 2.0\,r_0$, and $\phi_{M0} = 3.36 \times 10^6$ erg cm$^{-2}$ s$^{-1}$, yielding $u_\infty = 434.6$ km s$^{-1}$. The HZ radius for this star is also shown in Figure 2.

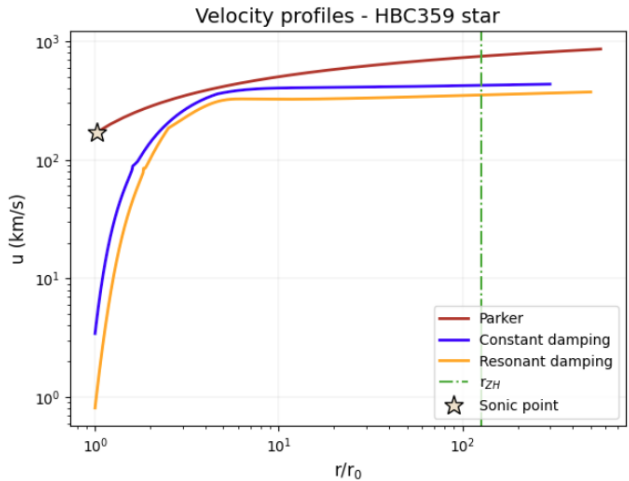

**Figure 2.** Velocity profiles of the HBC359 star considering the E. N. Parker (1958) and JPO89 models, with constant and resonant damping. It also shows the location of this star's HZ.

To estimate the magnetospheric size of the hypothetical Earth-like planets placed within the HZs of the WTTSs, we adopted a magnetic moment of $5.3 \times 10^{22}$ Am$^2$ (see Table 1), corresponding to the estimated magnetic moment of the Earth when the Sun was at an age comparable to those of the stars in our sample. Figures 3 and 4 show the relation between the magnetospheric sizes of the planets and the stellar masses of the 46 WTTSs, as well as their correlation with stellar chromospheric activity for both stellar-wind models. From the sample, 12 stars had double values of $L_X$, which are shown in the figures as a black vertical lines connecting both points. In Figure 3, they cannot be seen since their magnetospheric sizes were too close.

To estimate the uncertainties associated with the magnetospheric sizes derived from both the MHD-based stellar-wind models and the E. N. Parker (1958) wind model, we carried out a simple Monte Carlo analysis in which key input parameters were allowed to vary within physically motivated ranges. The adopted parameter ranges are intended to capture the large intrinsic dispersion and limited observational constraints characteristic of young stellar systems, such as the mass-loss rate (A. A. Vidotto & J. F. Donati 2017), the planetary magnetic moment (L. Ziegler et al. 2011), and the stellar surface magnetic field strength (C. Hill et al. 2019). For some other parameters, we adopted an error percentage associated with the calculated value, such as the stellar-wind velocity at the HZ ($u_{HZ}$), the coronal electron number density ($\tilde{n}_c$), and the radial location of the HZ ($r_{HZ}$), whose adopted value corresponds to the entire extent of the HZ (moist to maximum greenhouse). Table 2 summarizes the parameters varied in this analysis. The resulting magnetospheric size intervals therefore represent the central 68% of the distribution obtained from Monte Carlo sampling, corresponding to the range of physically plausible solutions within the adopted parameter space.

**Table 1**
Magnetic Moment Values for Different Ages of the Earth

| Era | Magnetic Moment (Am$^2$) |
|---|---|
| WTTS ages (∼2 Myr) | [a] $5.3 \times 10^{22}$ |
| Paleoarchean Earth (∼3.5 Gyr) | [b] $4.4 - 4.8 \times 10^{22}$ |
| Present-day Earth (∼5 Gyr) | [c] $8.0 \times 10^{22}$ |

**Notes.**
[a] L. Ziegler et al. (2011) and R. K. Bono et al. (2022).
[b] J. A. Tarduno et al. (2010), A. J. Biggin et al. (2015), and R. K. Bono et al. (2022).
[c] J. A. Tarduno et al. (2010).

It is expected that there is a 68% chance that the expected value of the standoff distance will be within the estimated range for each star. Figure 3 shows the results for the MHD-based model, where we can notice a large and asymmetrical interval of values, especially for constant damping. This happens due to the range of values adopted for the mass-loss rate, so that what we calculate is close to the lower limit, which is why it is shorter. Figure 4 represents the standoff distance for the E. N. Parker (1958) model. In this case, we can see a more symmetrical representation of the intervals, due to the fixed value in the percentage error of the coronal electron number density. It also includes, for reference, the estimated range of terrestrial magnetospheric sizes during the Paleoarchean era. During that time, Earth's magnetic moment is estimated to have ranged between 4.4 and $4.8 \times 10^{22}$ Am$^2$, when the Earth was 2400–3450 Myr. Table 1 shows the values of the Earth's magnetic moment according to its age.

Based on our adopted physical conditions, a magnetospheric size of $5\,R_?$ is used as a reference threshold, corresponding to estimates for the Earth during the Paleoarchean period. Within this framework, planetary magnetospheres exceeding $5\,R_?$ correspond to cases in which the planetary magnetic pressure is sufficient to balance the stellar-wind ram and magnetic pressures at larger standoff distances, whereas smaller magnetospheres indicate systems where the stellar wind compresses the magnetosphere more efficiently during the early, highly active phase of stellar evolution. Overall, our results show that stars at this early evolutionary stage predominantly produce small planetary magnetospheres across all wind models considered, with only a limited number of cases exceeding the Paleoarchean limit, as illustrated in Figures 3 and 4.

Figure 5 presents, for comparison, the planetary magnetospheric sizes for the E. N. Parker (1958) model, which we calculated as a function of stellar age for 41 WTTSs, and 120 solar-like stars calculated by V. See et al. (2014), since age estimates were not available for the entire sample. For the WTTSs, M. Güdel et al. (2007) calculated the ages using isochrones, and for solar-like stars, by their chromospheric activity (S. Marsden et al. 2014). Additionally, a clear correlation is observed between chromospheric activity and magnetospheric size in both samples. While WTTSs are expected to exhibit higher levels of chromospheric activity, our results do not show this trend. This discrepancy is likely due to the use of a different chromospheric activity model than that employed by V. See et al. (2014). We also note that in our analysis the planetary magnetic moment was scaled with stellar age, while V. See et al. (2014) assumed a fixed value over the entire 0.3–10 Gyr range. Figure 5 also includes the present-day size of the Earth's magnetosphere ($10.2\,R_\oplus$), assuming a current magnetic moment of $8.0 \times 10^{22}\,\mathrm{Am}^2$. Considering the full combined sample, spanning stellar ages from ∼0.5 Myr to 10 Gyr, we find a clear relationship between stellar age and planetary magnetospheric size, with very young stars systematically associated with smaller magnetospheres.

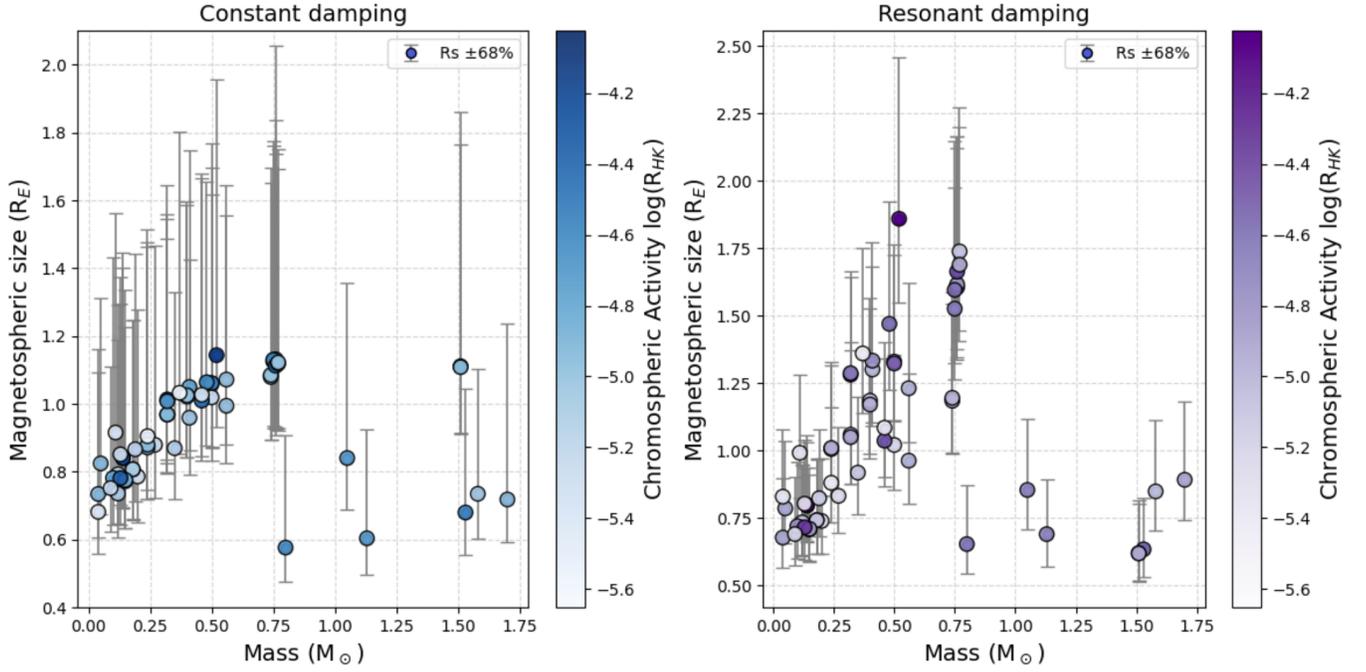

**Figure 3.** Relation between the planetary magnetosphere size and the stellar mass for the JPO89 model with constant (left) and resonant (right) damping (see Table A1), color coded by chromospheric activity.

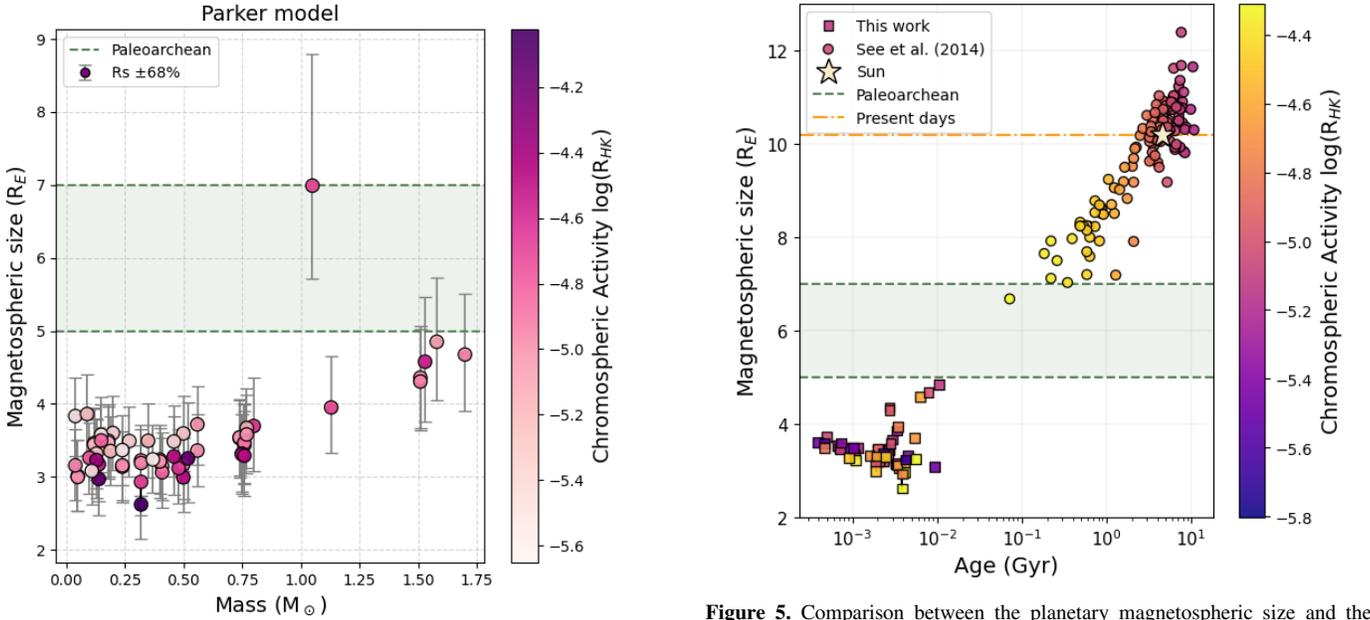

**Figure 4.** Same relation as Figure 3, but for the E. N. Parker (1958) wind model. It also shows the magnetospheric size of the Earth during the Paleoarchean period.

**Figure 5.** Comparison between the planetary magnetospheric size and the star's age for the E. N. Parker (1958) model using WTTS data and data from V. See et al. (2014), color coded by chromospheric activity. The figure also includes the Earth's position, based on the Sun, which has an age of 4.57 Gyr (A. Bonanno et al. 2002).

The magnetic field topology of PMS stars strongly depend on their internal structure. As stars develop a radiative core, the dipolar component weakens, reflecting the transition from fully to partially convective regimes. As PMS stars contract and evolve, both stellar mass and age play key roles in shaping their magnetic field properties (S. Gregory et al. 2012). Observations also reveal differences in X-ray emission, where PMS stars on radiative tracks show systematically lower $\log(L_X/L_{\rm bol})$ compared to fully convective ones (S. G. Gregory et al. 2016). Figure 6 illustrates the relation between the Ca II magnetospheric size. It was expected that those with larger magnetospheres would be in the lower branch. N. Astudillo-Defru et al. (2017) show that activity level probably decreases with stellar mass for M dwarfs, which have an internal structure similar to that of WTTSs (J. Morin et al. 2008; J.-F. Donati et al. 2010). Also, they found that when comparing chromospheric activity with stellar mass, the distribution is approximately flat above $\sim 0.6\,M_\odot$, and flattens again below $\sim 0.35\,M_\odot$, where stars with $M_\star < 0.35\,M_\odot$ are expected to be fully convective stars, which may reflect a transition at $\sim 0.35\,M_\odot$, associated with changes in both the dynamo and the large-scale magnetic field topology. We can see chromospheric activity increasing in $0.35\,M_\odot < M_\star < 0.6\,M_\odot$, flat behavior for $M_\star > 0.6\,M_\odot$, and some dispersion for $M_\star < 0.35\,M_\odot$.

## 5. Summary and Conclusions

We investigated the interaction between stellar winds from WTTSs and the magnetospheres of hypothetical Earth-like planets placed within the circumstellar HZs of their host stars. For this analysis, we placed a hypothetical terrestrial planet within the HZ of each host star and computed the pressure balance between the stellar wind and the planetary magnetosphere at that location. This calculation was performed using two distinct stellar-wind models: the E. N. Parker (1958) model and the JPO89 model, incorporating both constant and resonant damping of Alfvén waves. The resulting magnetospheric sizes provide a quantitative measure of how strongly planetary magnetic fields are compressed by stellar winds in very young stellar environments.

WTTSs represent an extreme regime in terms of stellar-wind conditions, characterized by high levels of magnetic activity and strong wind pressures. As a consequence, planetary magnetospheres around such stars are generally small across all wind models considered, and planets orbiting them are particularly susceptible to atmospheric erosion. To investigate how magnetospheric size evolves with stellar age, we followed the approach of V. See et al. (2014) and observed that magnetospheric protection provided by a fixed planetary magnetic field can increase over time, as a consequence of the declining stellar magnetic activity. As stars age and spin down, their chromospheric and overall magnetic activity decrease, allowing planetary magnetospheres to expand and provide more effective atmospheric shielding. We also considered that some of our WTTSs may be fully convective, while others may already have developed a small radiative core. This structural difference can directly influence the behavior of their chromospheric activity.

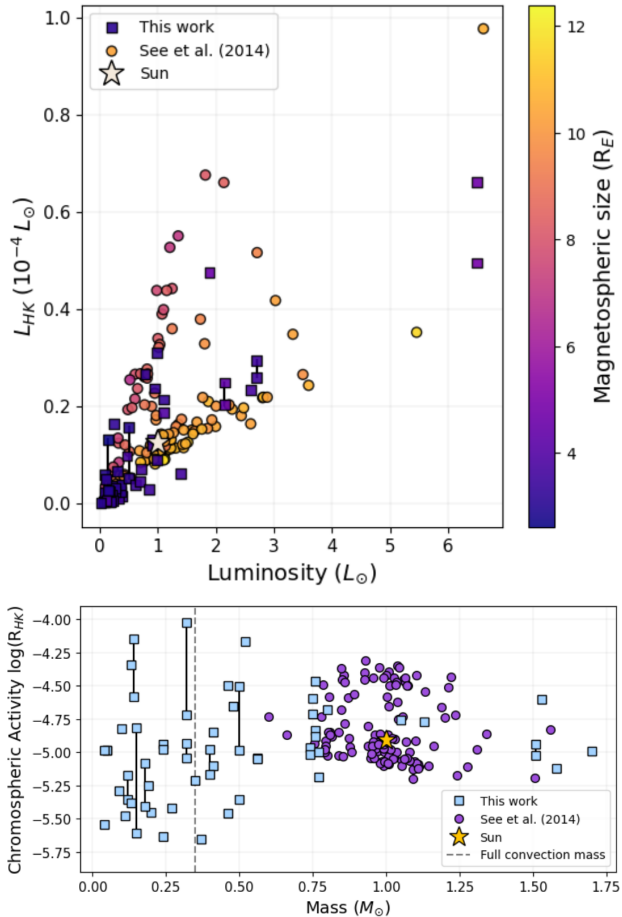

**Figure 6.** Upper panel: Ca II H and K line luminosities vs. bolometric luminosity, color coded by planetary magnetosphere size for the E. N. Parker (1958) model. Lower panel: chromospheric activity as a function of stellar mass, where the dashed line represents the fully convective limit of $0.35\,M_\odot$. Both panels consider WTTS data and solar-type stars from V. See et al. (2014).

**Table 2**
Parameters Varied in the Monte Carlo Uncertainty Analysis

| Parameter | Range or Value |
| --- | --- |
| $\dot{M}$ ($M_\odot\,\mathrm{yr}^{-1}$) | $10^{-12}$–$10^{-9}$ |
| $\mathcal{M}$ (Am$^2$) | $\pm 1.5 \times 10^{22}$ |
| $r_{\mathrm{HZ}}$ (%) | 25 |
| $u_{\mathrm{HZ}}$ (%) | 30 |
| $B_0$ (G) | 500–2000 |
| $\tilde{n}_c$ (%) | 95 |

H and K line luminosities and the bolometric luminosity, highlighting their connection with magnetospheric size. The figure also shows how chromospheric activity varies with stellar mass, both for our WTTS sample and for the solar-type stars from V. See et al. (2014). As noted previously by V. See et al. (2014), the Vaughan–Preston gap, defined as the separation between the two branches of activity, is already evident in their sample, for which different explanations have been proposed regarding its origin. By adding our WTTSs to this diagram, we find that most do not fall within the branches of the gap, while those that do are located either on the upper or lower branches and seem to have no relation to

## Acknowledgments

Y.F.T. acknowledges CAPES/PROEX (proc. 88887.820782/2023-00) and FAPESP (proc. 2021/11689-6). Y.F.T. also thanks Dr. Jane Gregorio-Hetem for discussions that contributed to improving the quality of this work. V.J.P. acknowledges FAPESP grant (proc. 2021/02120-0). N.F.S.A. acknowledges FAPESP (proc. 2019/26787-3).

## Appendix
### Additional Table

Table A1 summarizes the main properties of the stars and hypothetical planets analyzed in this work.

Table A1
Planetary Magnetospheres Sizes $R_s \pm 68\%$ of Hypothetical Earth-like Planets Centered in the Habitable Zone, $r_{HZ}$, of Weak-lined T Tauri Stars for the E. N. Parker (1958) and V. Jatenco-Pereira & R. Opher (1989a) Models, with Constant and Resonant Damping, Where We Adopt an Initial Damping Length of $L_0 = 4.0\, r_0$, Except for Those Marked with an "a" Based on Damping (Columns 5 and 6).

| Star ID | $L_{X\star}$ ($10^{30}$) (erg s$^{-1}$) | Age (Myr) | $r_{HZ}$ (au) | $R_{s,JPO89}^{constant}$ ($R_\oplus$) | $R_{s,JPO89}^{resonant}$ ($R_\oplus$) | $R_s^{Parker}$ ($R_\oplus$) |
|---|---|---|---|---|---|---|
| HBC 352 | 2.657 | ... | 1.13 | $0.84^{+0.52}_{-0.15}$ | $0.85^{+0.26}_{-0.15}$ | $7.02^{+1.82}_{-1.25}$ |
| HBC 358 AB | 0.383 | 3.26 | 0.81 | $0.96^{+0.62}_{-0.17}$ | $1.30^{+0.38}_{-0.21}$ | $3.13^{+0.47}_{-0.46}$ |
| HBC 359 | 0.663 | 3.54 | 0.79 | $^a1.04^{+0.61}_{-0.18}$ | $1.33^{+0.44}_{-0.23}$ | $3.05^{+0.49}_{-0.47}$ |
| LkCa 1 | 0.232 | 0.87 | 0.96 | $0.88^{+0.54}_{-0.15}$ | $0.83^{+0.26}_{-0.14}$ | $3.46^{+0.49}_{-0.47}$ |
| Anon 1 | 4.139 | 0.50 | 2.45 | $0.99^{+0.54}_{-0.17}$ | $0.96^{+0.32}_{-0.16}$ | $3.73^{+0.52}_{-0.56}$ |
| V773 Tau ABC | 9.488 | 6.35 | 1.96 | $0.68^{+0.41}_{-0.12}$ | $0.63^{+0.19}_{-0.10}$ | $4.57^{+0.94}_{-0.80}$ |
| CIDA 2 | 0.178 | 0.39 | 0.89 | $^a0.79^{+0.46}_{-0.14}$ | $0.74^{+0.23}_{-0.12}$ | $3.60^{+0.53}_{-0.51}$ |
| LkCa 5 | 0.432 | 2.31 | 0.94 | $1.02^{+0.48}_{-0.17}$ | $1.18^{+0.38}_{-0.20}$ | $3.25^{+0.47}_{-0.47}$ |
| LkCa 5 | 0.681 | 2.31 | 0.94 | $1.02^{+0.49}_{-0.18}$ | $1.17^{+0.36}_{-0.20}$ | $3.20^{+0.49}_{-0.47}$ |
| CIDA 3 | 0.277 | 3.34 | 0.58 | $0.87^{+0.48}_{-0.15}$ | $1.01^{+0.32}_{-0.17}$ | $3.14^{+0.45}_{-0.48}$ |
| CIDA 3 | 0.254 | 3.34 | 0.58 | $0.88^{+0.57}_{-0.16}$ | $1.01^{+0.31}_{-0.18}$ | $3.16^{+0.46}_{-0.48}$ |
| V410 X3 | 0.059 | 2.68 | 0.46 | $0.79^{+0.54}_{-0.14}$ | $0.73^{+0.22}_{-0.12}$ | $3.45^{+0.49}_{-0.49}$ |
| V410 X3 | 0.092 | 2.68 | 0.46 | $0.73^{+0.38}_{-0.13}$ | $0.73^{+0.21}_{-0.12}$ | $3.45^{+0.48}_{-0.49}$ |
| V410 Tau ABC | 3.762 | 2.74 | 2.13 | $1.12^{+0.85}_{-0.20}$ | $0.62^{+0.19}_{-0.10}$ | $4.36^{+0.70}_{-0.68}$ |
| V410 Tau ABC | 4.663 | 2.74 | 2.13 | $1.11^{+0.55}_{-0.20}$ | $0.62^{+0.18}_{-0.10}$ | $4.30^{+0.70}_{-0.64}$ |
| CZ Tau AB | 0.423 | 2.10 | 0.81 | $0.97^{+0.63}_{-0.17}$ | $1.06^{+0.34}_{-0.19}$ | $3.23^{+0.49}_{-0.48}$ |
| CZ Tau AB | 0.547 | 2.10 | 0.81 | $0.97^{+0.57}_{-0.17}$ | $1.05^{+0.32}_{-0.17}$ | $3.19^{+0.49}_{-0.46}$ |
| V410 X7 | 0.913 | 1.90 | 1.08 | $1.06^{+0.64}_{-0.18}$ | $1.33^{+0.43}_{-0.22}$ | $3.19^{+0.50}_{-0.49}$ |
| V410 X7 | 2.975 | 1.90 | 1.08 | $1.07^{+0.69}_{-0.18}$ | $1.32^{+0.40}_{-0.22}$ | $3.00^{+0.53}_{-0.49}$ |
| Hubble 4 | 5.342 | 0.68 | 2.47 | $1.08^{+0.57}_{-0.19}$ | $1.18^{+0.36}_{-0.19}$ | $3.52^{+0.51}_{-0.51}$ |
| Hubble 4 | 4.668 | 0.68 | 2.47 | $1.08^{+0.62}_{-0.20}$ | $1.19^{+0.39}_{-0.21}$ | $3.55^{+0.52}_{-0.54}$ |
| KPNO-Tau 2 | 0.012 | ... | 0.13 | $0.82^{+0.50}_{-0.14}$ | $0.78^{+0.25}_{-0.13}$ | $3.02^{+0.47}_{-0.48}$ |
| V410 X6 | 0.124 | 0.71 | 0.70 | $0.81^{+0.47}_{-0.15}$ | $0.74^{+0.22}_{-0.12}$ | $3.48^{+0.50}_{-0.47}$ |
| V410 X6 | 0.278 | 0.71 | 0.70 | $0.81^{+0.48}_{-0.15}$ | $0.74^{+0.23}_{-0.13}$ | $3.47^{+0.48}_{-0.49}$ |
| V410 X5 | 0.387 | 4.21 | 0.45 | $0.84^{+0.54}_{-0.14}$ | $0.80^{+0.24}_{-0.14}$ | $3.15^{+0.50}_{-0.50}$ |
| V410 X5 | 1.115 | 4.21 | 0.45 | $0.84^{+0.54}_{-0.15}$ | $0.80^{+0.26}_{-0.14}$ | $2.97^{+0.57}_{-0.51}$ |
| V819 Tau AB | 2.445 | 2.02 | 1.43 | $1.12^{+0.62}_{-0.20}$ | $1.60^{+0.52}_{-0.27}$ | $3.42^{+0.53}_{-0.54}$ |
| V819 Tau AB | 2.205 | 2.02 | 1.43 | $1.12^{+0.62}_{-0.20}$ | $1.62^{+0.55}_{-0.27}$ | $3.46^{+0.56}_{-0.56}$ |
| 2M J04213459 | 0.043 | 4.56 | 0.40 | $0.85^{+0.43}_{-0.15}$ | $0.80^{+0.24}_{-0.13}$ | $3.32^{+0.46}_{-0.45}$ |
| HD 283572 | 13.003 | 7.92 | 3.42 | $0.71^{+0.43}_{-0.12}$ | $0.89^{+0.29}_{-0.15}$ | $4.69^{+0.83}_{-0.78}$ |
| LkCa 21 | 0.646 | 1.16 | 1.22 | $0.87^{+0.53}_{-0.15}$ | $0.92^{+0.28}_{-0.15}$ | $3.50^{+0.49}_{-0.51}$ |
| KPNO-Tau 13 | 0.138 | 2.66 | 0.61 | $0.86^{+0.62}_{-0.15}$ | $0.82^{+0.25}_{-0.14}$ | $3.35^{+0.48}_{-0.50}$ |
| DI Tau AB | 1.568 | 1.07 | 1.51 | $1.08^{+0.77}_{-0.19}$ | $1.23^{+0.39}_{-0.20}$ | $3.38^{+0.50}_{-0.49}$ |
| KPNO-Tau 5 | 0.010 | ... | 0.24 | $0.68^{+0.35}_{-0.12}$ | $0.83^{+0.25}_{-0.14}$ | $3.80^{+0.56}_{-0.53}$ |
| MHO 9 | 0.080 | 2.33 | 0.73 | $^a0.90^{+0.73}_{-0.16}$ | $0.88^{+0.28}_{-0.15}$ | $3.38^{+0.46}_{-0.48}$ |
| MHO 4 | 0.123 | ... | 0.35 | $0.78^{+0.42}_{-0.14}$ | $0.72^{+0.24}_{-0.12}$ | $3.27^{+0.48}_{-0.49}$ |
| L1551 51 | 1.841 | 5.42 | 1.03 | $0.58^{+0.29}_{-0.10}$ | $0.65^{+0.22}_{-0.11}$ | $3.69^{+0.67}_{-0.60}$ |
| V827 Tau | 4.010 | 1.59 | 1.57 | $1.12^{+0.57}_{-0.20}$ | $1.53^{+0.45}_{-0.26}$ | $3.30^{+0.53}_{-0.53}$ |
| V826 Tau | 4.523 | 1.94 | 1.46 | $^a1.12^{+0.62}_{-0.20}$ | $1.60^{+0.55}_{-0.27}$ | $3.29^{+0.59}_{-0.54}$ |
| V928 Tau AB | 1.046 | 0.73 | 1.80 | $^a1.02^{+0.64}_{-0.18}$ | $1.02^{+0.33}_{-0.17}$ | $3.59^{+0.51}_{-0.52}$ |
| MHO 8 | 0.065 | 0.47 | 0.65 | $^a0.78^{+0.45}_{-0.14}$ | $0.71^{+0.20}_{-0.12}$ | $3.57^{+0.49}_{-0.50}$ |
| MHO 8 | 0.444 | 0.47 | 0.65 | $^a0.77^{+0.39}_{-0.13}$ | $0.71^{+0.23}_{-0.12}$ | $3.50^{+0.52}_{-0.52}$ |
| KPNO-Tau 14 | 0.923 | 1.09 | 0.52 | $^a0.78^{+0.60}_{-0.14}$ | $0.72^{+0.22}_{-0.12}$ | $3.25^{+0.54}_{-0.51}$ |
| V830 Tau | 5.181 | 2.48 | 1.33 | $1.11^{+0.68}_{-0.20}$ | $1.66^{+0.50}_{-0.29}$ | $3.30^{+0.59}_{-0.54}$ |
| JH 108 | 1.233 | 3.40 | 0.84 | $^a1.06^{+0.61}_{-0.18}$ | $1.47^{+0.45}_{-0.24}$ | $3.11^{+0.55}_{-0.50}$ |
| FF Tau AB | 0.796 | 2.95 | 1.25 | $1.12^{+0.71}_{-0.20}$ | $1.74^{+0.53}_{-0.29}$ | $3.67^{+0.57}_{-0.54}$ |
| CoKu Tau 3 AB | 5.851 | 0.92 | 1.51 | $1.01^{+0.62}_{-0.19}$ | $1.04^{+0.31}_{-0.17}$ | $3.28^{+0.55}_{-0.53}$ |
| KPNO-Tau 15 | 2.624 | 3.86 | 0.58 | $1.01^{+0.73}_{-0.18}$ | $1.28^{+0.36}_{-0.21}$ | $2.63^{+0.56}_{-0.47}$ |
| KPNO-Tau 15 | 0.483 | 3.86 | 0.58 | $1.01^{+0.56}_{-0.17}$ | $1.29^{+0.38}_{-0.22}$ | $2.92^{+0.47}_{-0.45}$ |
| HP Tau/G3 AB | 1.293 | 2.83 | 1.27 | $1.13^{+0.67}_{-0.20}$ | $1.69^{+0.51}_{-0.28}$ | $3.60^{+0.55}_{-0.55}$ |
| HP Tau/G2 | 9.653 | 10.5 | 3.36 | $0.74^{+0.41}_{-0.13}$ | $0.85^{+0.26}_{-0.14}$ | $4.81^{+0.89}_{-0.76}$ |
| CFHT-BD Tau 3 | 0.012 | ... | 0.13 | $^a0.74^{+0.42}_{-0.13}$ | $0.68^{+0.22}_{-0.11}$ | $3.18^{+0.50}_{-0.48}$ |
| JH 223 | 0.064 | 4.27 | 0.65 | $1.03^{+0.70}_{-0.18}$ | $1.36^{+0.39}_{-0.22}$ | $3.24^{+0.46}_{-0.46}$ |
| Haro 6-32 | 0.101 | 3.33 | 0.54 | $^a0.76^{+0.44}_{-0.13}$ | $0.69^{+0.21}_{-0.12}$ | $3.86^{+0.54}_{-0.56}$ |
| CoKuLk332/G2 AB | 3.257 | 5.60 | 0.75 | $1.14^{+0.80}_{-0.21}$ | $^a1.86^{+0.60}_{-0.31}$ | $3.22^{+0.77}_{-0.56}$ |

**Table A1**
(Continued)

| Star ID | $L_{X\star}$ ($10^{30}$) (erg s$^{-1}$) | Age (Myr) | $r_{HZ}$ (au) | $R_{s,JPO89}^{constant}$ ($R_?$) | $R_{s,JPO89}^{resonant}$ ($R_?$) | $R_s^{Parker}$ ($R_?$) |
|---|---|---|---|---|---|---|
| CoKuLk332/G1 AB | 0.493 | 1.02 | 1.41 | $1.03^{+0.79}_{-0.19}$ | $1.08^{+0.32}_{-0.18}$ | $3.48^{+0.49}_{-0.50}$ |
| 2M J04554046+30 | 0.011 | 9.39 | 0.23 | $0.92^{+0.50}_{-0.16}$ | $0.99^{+0.44}_{-0.17}$ | $3.07^{+0.44}_{-0.44}$ |
| HBC 427 | 3.558 | 3.43 | 1.55 | $0.60^{+0.38}_{-0.11}$ | $0.69^{+0.20}_{-0.12}$ | $3.99^{+0.68}_{-0.68}$ |

**Notes.** It also shows their X-ray luminosity and age (M. Güdel et al. 2007).
[a] Initial damping length adopted as $L_0 = 2.0\, r_0$.

### ORCID iDs

Yasmmin F. Tamburus  https://orcid.org/0000-0003-4682-2459
Natália F. S. Andrade  https://orcid.org/0000-0001-5970-334X
Guilherme R. C. Sampaio  https://orcid.org/0009-0008-7350-7546
Vera Jatenco-Pereira  https://orcid.org/0000-0002-1517-0710